\documentclass[runningheads]{llncs}

\usepackage{eccv}

\usepackage{eccvabbrv}

\usepackage{graphicx}
\usepackage{booktabs}
\usepackage{tabularx}
\usepackage{array}

\usepackage[accsupp]{axessibility}  

\usepackage{hyperref}

\usepackage{orcidlink}

\begin{document}

\title{Evaluating Usability and Engagement of Large Language Models in Virtual Reality for Traditional Scottish Curling} 

\titlerunning{Usability and Engagement of Large Language Models in Virtual Reality}

\author{Ka Hei Carrie Lau\inst{1}\orcidlink{0009-0005-8838-3230} \and
Efe Bozkir\inst{1}\orcidlink{0000-0002-4594-4318} \and
Hong Gao\inst{1}\orcidlink{0000-0003-3934-433X} \and
Enkelejda Kasneci\inst{1}\orcidlink{0000-0003-3146-4484}}

\authorrunning{Lau et al.}

\institute{Human-Centered Technologies for Learning, Technical University of Munich, Munich, Germany
\email{\{carrie.lau, efe.bozkir, hong.gao, enkelejda.kasneci\}@tum.de}}

\maketitle

\begin{abstract}

This paper explores the innovative application of Large Language Models (LLMs) in Virtual Reality (VR) environments to promote heritage education, focusing on traditional Scottish curling presented in the game ``Scottish Bonspiel VR''. Our study compares the effectiveness of LLM-based chatbots with pre-defined scripted chatbots, evaluating key criteria such as usability, user engagement, and learning outcomes. The results show that LLM-based chatbots significantly improve interactivity and engagement, creating a more dynamic and immersive learning environment. This integration helps document and preserve cultural heritage and enhances dissemination processes, which are crucial for safeguarding intangible cultural heritage (ICH) amid environmental changes. Furthermore, the study highlights the potential of novel technologies in education to provide immersive experiences that foster a deeper appreciation of cultural heritage. These findings support the wider application of LLMs and VR in cultural education to address global challenges and promote sustainable practices to preserve and enhance cultural heritage.

\end{abstract}

\section{Introduction}
In recent years, integrating digital technologies in education has transformed learning paradigms, making them more immersive, interactive, and personalized~\cite{HALEEM2022275}. Virtual Reality (VR), in particular, has demonstrated its capacity to create highly immersive environments that simulate real-life experiences or historical events~\cite{yildirim2018analysis}. Similarly, advancements in Large Language Models (LLMs) enable the generation of human-like responses, providing interactive learning experiences. However, combining VR and LLMs to preserve cultural heritage remains unexplored. Specifically, enhancing cultural learning by immersing learners in virtual environments enriched with intelligent interactions presents a unique opportunity~\cite{bozkir2024embedding_arxiv}. This paper aims to evaluate user perception and the educational impact of integrating these technologies in virtual environments for cultural learning. This area requires further research and presents a promising direction for future studies.

Global challenges, such as climate change, threaten cultural heritage, impacting both tangible and intangible aspects vital to human well-being and cultural identity~\cite{Higgins2022, Pearson2023}. For instance, the historic ``Grand Match'', a curling competition reliant on natural ice since 1847, faced discontinuation after 1979 due to unstable ice conditions caused by climate change~\cite{2010_GrandMatch_Safety}. This highlights the urgent need to preserve cultural practices vulnerable to environmental changes and explores how modern technologies can aid this endeavor.

Currently, cultural preservation often relies on institutions like museums and archives, which struggle to engage younger and digitally-oriented audiences~\cite{Addis2005NewTA, Vasile2015}. While cultural institutions have adopted technologies like VR and Natural Language Processing (NLP) chatbots, challenges remain, such as limited conversational depth in chatbots~\cite{Varitimiadis2020, Gaia2019} and fragmented narrative delivery in VR~\cite{Hudson2018, INNOCENTE2023268}. These issues indicate an insufficient research in effectively integrating technology into cultural education.

Our study addresses these challenges by exploring the integration of LLMs into VR environments, comparing LLM-based solutions like ChatGPT with predefined chatbots to enhance usability and engagement in cultural education. We developed the \textit{Scottish Bonspiel VR} prototype, leveraging multimodal interaction to offer holistic cultural heritage experiences and highlight the environmental impact on traditions like the Grand Match. This investigation addresses a critical research gap at the intersection of LLMs and VR, exploring which set-up most effectively enhances user engagement and usability, contributing to the broader discourse on integrating emerging technologies into cultural and educational frameworks.

Experimental Preservation is a concept that challenges traditional cultural heritage practices by emphasizing inclusivity and innovative methodologies to preserve overlooked objects and sites, making cultural heritage more relevant to contemporary society~\cite{Otero-Pailos2016-xv}. By integrating VR and LLM technologies, our research aligns with this approach, creating dynamic and interactive preservation experiences that engage diverse audiences.

Our research advances the state of the art with the following contributions:
\begin{enumerate}
\item Development of an interactive system combining LLMs and VR to enhance heritage education accessibility and improve the learning experience.
\item An exploratory user study evaluating usability, user engagement, learning outcomes, and cognitive load with an LLM-based chatbot in a VR environment for cultural heritage preservation and education.
\item Identification of usability aspects such as accuracy, intuitiveness, communication, overall experience, response time, and satisfaction of LLM-based chatbots as cultural heritage ambassadors.
\end{enumerate}

In this study, we aim to bridge the gap between the potential of emerging technologies, such as VR and LLMs, and their practical application in developing interactive systems for effective cultural heritage education.

\section{Related Work}
We provide an overview of the previous work in two directions, which are discussed in the following subsections.

\subsection{Preserving Cultural Heritage in Digital Humanities}
Cultural spaces have increasingly adopted digital technologies such as digital storytelling~\cite{Pujol2011PersonalizingID, Hupperetz2012, Liestøl2014}, serious games~\cite{Paliokas2016, Anderson2010}, and interactive displays~\cite{Hornecker2008, Reunanen2015} to engage users. VR, in particular, has shown promise in Digital Humanities (DH) for enhancing cultural heritage experiences by enriching user interactions and improving educational outcomes~\cite{Marsili_2020, Lugmayr_2015, Hutson_Olsen_2022, Bekele2018}. However, these implementations often focus on individual tools rather than the holistic integration of technologies.

Our study explores this space by merging VR and LLMs to improve cultural heritage accessibility and interactivity. By leveraging LLM chatbots and web-based VR environments, we aim to enhance dynamic user engagement beyond that provided by traditional conversational chatbots~\cite{Varitimiadis2020} and audio guides~\cite{Gaia2003}. This approach seeks to document, preserve, and disseminate cultural heritage while evaluating system usability, engagement, and learning outcomes, thereby contributing to the preservation of cultural heritage in the DH field.

\subsection{The Potentials of Emerging Technologies on New Learning Paradigms}
\subsubsection{Towards Human-like Conversational Chatbots for Cultural Heritage}
LLM-based chatbots have significantly improved the delivery of historical content in museums by leveraging NLP and ML for enhanced visitor interactions~\cite{Varitimiadis2020, Gaia2019}. However, these technological solutions often face scalability challenges beyond the confines of physical institutions, which restricts their broader applicability on global cultural heritage education.~\cite{Damala2019, Boiano2018, Gaia2019}. Despite interactive and personalized recommendations~\cite{Boiano2016, Casillo2022}, challenges in achieving deep conversational contexts persist~\cite{Gaia2019, Casillo2022, Colace2020}.

Recent advancements in Human-Centered LLM, particularly through explainable LLM (XLLM)~\cite{Mohseni2021} and GPT-4~\cite{openai2023gpt4}, promise new horizons for cultural heritage accessibility~\cite{Pisoni2021}. XLLM strategies deepen user engagement through interactive storytelling and participatory narratives~\cite{Trichopoulos2023_Survey}. ChatGPT, for instance, offers comprehensive and contextually relevant interactions~\cite{yenduri2023generative, JEON2023104898}.

Our research integrates an LLM-based chatbot within a VR environment to scale cultural heritage experiences beyond museum spaces. This approach leverages LLM's advanced conversational abilities to provide more meaningful and accessible cultural heritage interactions.

\subsubsection{VR for Intangible Cultural Heritage (ICH) Preservation}
VR has revolutionized how historical and cultural content is experienced, enhancing visitor immersion and engagement~\cite{Trunfio2021, Duranti2024}. However, VR's potential for ICH preservation remains underexplored, even though ICH is more vulnerable to loss~\cite{Pătru-Stupariu2019, Higgins2022, Pearson2023}. VR can revive historical events and present ICH practices to a broader audience, overcoming geographical barriers~\cite{Hou2022}.

ICH, transmitted through oral traditions and social practices, poses documentation challenges~\cite{Higgins2022, Wulf2004CrucialPI}. VR's ability to create interactive spaces offers a novel approach to fostering community participation and understanding of ICH~\cite{Mah2019, Ch’ng2020, DEPAOLIS2022e00238}. It facilitates embodiment learning, allowing users to actively participate in cultural practices through simulation~\cite{Hou2022, Skovfoged2018, Zhang2023, Aristidou2021}.

However, VR's application in ICH preservation faces challenges such as potential loss of context, emotional depth~\cite{Idris2016} and scalability issues~\cite{Hou2022}. Our work introduces an interactive system that leverages web-based VR integrated with LLMs to address these limitations, making ICH more accessible and adaptable to evolving technologies. This approach seeks to preserve the richness of cultural heritage and enhance engagement beyond traditional museum settings.

\section{Methods}
This study aimed to enhance user experience within cultural heritage applications by integrating VR and LLMs. We developed the ``Scottish Bonspiel VR'' game, incorporating an LLM-based chatbot as a cultural ambassador to illustrate how traditional Scottish curling was played. This exploratory research compared the performance of ChatGPT with a predefined chatbot commonly used in digital cultural heritage settings. The predefined chatbot, featuring scripted responses within Scottish cultural heritage contexts, served as a baseline for comparison. Without further fine-tuning of ChatGPT and the prescripted chatbot, this setup allowed for a direct comparison with the dynamic capabilities of the LLM-based chatbot, highlighting potential improvements in usability, engagement, and learning outcomes. Through this approach, we aimed to advance the application of chatbot technology in cultural heritage settings.

The Institutional Review Board (IRB) of the Technical University of Munich approved our study, ensuring adherence to ethical research standards. Our research aims to answer the following research questions:
\begin{itemize}
    \item \textbf{RQ1}: How does the LLM chatbot in a Scottish Bonspiel VR experience improve usability compared to a predefined chatbot?
    \item \textbf{RQ2}: How does the LLM chatbot in a Scottish Bonspiel VR experience enhance user engagement compared to a predefined chatbot?
    \item \textbf{RQ3}: How does the LLM chatbot in a Scottish Bonspiel VR experience improve cultural heritage learning outcomes relative to a predefined chatbot?
\end{itemize}

We employed a between-subjects design to evaluate the user experiences of two different chatbot technologies within the ``Scottish Bonspiel VR'' game. This exploratory research method involved comparing participant interactions with an LLM-based chatbot against those with a conventional predefined chatbot, scripted based on the history of Scottish Curling. The following subsections provide detailed information on participants, apparatus, tools, and specifics of our experimental design, procedures, measures, and analyses.

\subsection{Participants}
We recruited 36 participants, each compensated with a 10 Euro Amazon voucher. Eligibility criteria included being at least 18 years old, having a normal or corrected-to-normal vision, and fluency in English. A pre-study questionnaire screened for severe motion sickness or adverse reactions to VR environments. We excluded susceptible individuals to ensure safety and research integrity.

Participants were academically diverse, including bachelor's, master's, and doctoral students, with 22 males (\(M_{\text{age}} = 26.27\), \(SD_{\text{age}} = 3.54\)) and 14 females (\(M_{\text{age}} = 25.64\), \(SD_{\text{age}} = 3\)). Participants were randomly assigned into two groups: 19 interacted with the ChatGPT (CGPT) chatbot, and 17 used the predefined (PD) chatbot.

An initial interest form assessed their familiarity with VR, gaming experience, and Scottish curling knowledge. This included questions like ``Have you previously experienced virtual reality technology?'' and ``How experienced are you with playing computer games?''.~\cref{tbl_1} in the appendix details the participant demographics, showing varied experiences with VR, gaming and prior awareness of Scottish curling.

According to the Honey and Mumford Learning Styles Questionnaire (LSQ), ~\cite{honey1986manual}, which identifies four types of learning styles, most of our participants are Reflectors, comprising 61.11\% of the total. More details can be found in the appendix~\cref{tbl_1}. Reflectors prefer to learn by observing and reflecting, influencing their interactions with the chatbot. The diversity in learning styles helped us evaluate how different approaches impact the effectiveness and adaptability of LLM-based chatbots, enhancing our understanding of human-artificial intelligence (AI) interactions and promoting effective active learning.

\subsection{Apparatus and Tools}
We used the Oculus Quest 2\footnote{\url{https://www.meta.com/de/en/quest/products/quest-2/}, last accessed on 27 April 2024} for the immersive VR experience. The VR scenes were self-designed and developed using Vite\footnote{\url{https://vitejs.dev/}, last accessed on 27 April 2024} and A-frame\footnote{\url{https://aframe.io/}, last accessed on 27 April 2024}, hosted on Vercel\footnote{\url{https://vercel.com/}, last accessed on 27 April 2024}.

We employed the OpenLLM API for the interactive chatbot, utilizing the GPT-4 model~\cite{openai2023gpt4} for its advanced conversational capabilities. To allow participants to communicate verbally with our chatbot, we integrated the OpenLLM API with Whisper\footnote{\url{https://openai.com/research/whisper}, last accessed on 27 April 2024} for Speech-to-Text (STT) functionality. Chatbot responses were relayed to participants using Azure Text-to-Speech (TTS)\footnote{\url{https://azure.microsoft.com/en-us/free/ai-services/search}, last accessed on 27 April 2024}, facilitating interactive communication in the virtual environment.

\subsection{Experimental Design}
Our study employed a between-subjects experimental design, with participants randomly allocated into two groups: a control group interacting with a predefined chatbot and a treatment group engaging with ChatGPT within a VR environment themed around Scottish curling. The primary independent variable was the type of chatbot: predefined responses versus ChatGPT-generated responses. The dependent variables included usability feedback (measured by accuracy, communication, intuitiveness, overall experience, response time, and satisfaction), user engagement, learning outcomes, and cognitive task load. Control measures included standardizing VR content across two game levels, maintaining uniform (Q\&A) sequences, and designing a fixed task order to ensure a consistent participant experience.

The experimental setup included two levels showcasing Scottish curling's adaptation to climate change:
\begin{itemize}
    \item \textbf{Level 1 ``Traditional Scottish Curling on a Frozen Lake''}: Participants experienced an outdoor VR environment simulating the Grand Match, learning about the sport's history, cultural significance, and environmental impact, and virtually shooting a curling stone on a frozen lake.
    \item \textbf{Level 2 ``Indoor Scottish Curling''}: This level highlighted the sport's shift to indoor settings due to climate change. Participants engaged in a Q\&A session with either ChatGPT or a predefined chatbot, building on Level 1 to deepen their understanding.
\end{itemize}

~\cref{tbl_2} and~\cref{tbl_3} in the appendix present the chatbot dialogues across two levels, focused on three educational themes: Scottish curling history, climate change, and modern adaptations. This study explored the effectiveness of ChatGPT versus predefined chatbots in enhancing usability, engagement, and learning outcomes within VR environments. By evaluating metrics like accuracy, intuitiveness, communication, overall experience, response time, and satisfaction, we aimed to assess how effectively LLMs in VR contribute to cultural heritage preservation.

\subsubsection{Scottish Bonspiel VR}
\textit{The VR Experience Game Background} is inspired by the last outdoor Scottish curling Grand Match in 1979. We designed the scene, avatar uniforms, and curling stone based on heritage artifacts referenced from a book on curling history~\cite{Smith_David_Curling}. As mentioned previously, the gameplay consists of two levels, as depicted in the appendix~\cref{fig:fig_player}.

\textit{The Chatbot as a Scottish Cultural Heritage Ambassador} was created for our VR experience to educate players about Scotland's cultural heritage, as shown in the appendix~\cref{fig_all_artifacts}. Dressed in an outfit inspired by the grand match era, the cartoon-style avatar avoids the uncanny valley effect highlighted by Chalmers et al.~\cite{Chalmers_2021} and enhances historical authenticity. Through interaction with the chatbot, players can learn about different cultural aspects of Scottish curling.

\textit{The VR Experience Game Mechanic} immerses participants in curling, reflecting the sport's evolution from outdoor to indoor settings. The game simplifies curling by allowing players to aim and shoot using a single button, with the goal of guiding the stone to the target, or ``the house.'' Visual cues, such as a red charging bar for shot power and a green stripe for optimal release, help players time their shots. The aiming direction is controlled by the player's head position, tracked by the VR headset and represented by a crosshair, offering an intuitive experience.

\subsection{Procedure}
Upon arrival, participants received a detailed briefing on the research objectives to ensure they understood the study's purpose. They were introduced to the VR setup and given basic operational instructions with the controller. Initially, participants completed the LSQ and answered questions about their knowledge of cultural heritage and climate change related to traditional outdoor sports.

Participants then engaged in the two-game levels of the VR experiment, the whole VR session lasting approximately 10 to 20 minutes. They were reminded to pause or exit if they felt uncomfortable or disoriented.

After the VR sessions, participants completed questionnaires to assess their understanding of cultural heritage concepts, usability, and the NASA Task Load Index (NASA TLX). All participants provided digital informed consent before participating. The entire process, from orientation to completing questionnaires, typically took about 40 to 45 minutes per participant. A concise summary of the procedure and experimental details is presented in the appendix~\cref{fig_2}.

\subsection{Measures}
We employed a mixed-methods approach to collect data, focusing on usability, user engagement, and learning outcomes to assess the integration of LLMs with VR for cultural heritage education. Our goal was to evaluate the system's efficacy in autonomously conveying cultural heritage content, particularly in situations lacking direct access to heritage sites or experts.

\textit{Usability} was addressed through a customized usability survey based on the System Usability Scale (SUS)~\cite{Webster_2017} and the AttrakDiff questionnaire~\cite{Hassenzahl2003}. Details of the questionnaire can be found in the appendix~\cref{tbl_usability}. Participants rated their experience on a Likert scale from 1 (not at all) to 7 (very much), assessing dimensions such as accuracy, communication, intuitiveness, overall experience, response time, and satisfaction. An open-ended section collected qualitative feedback on technical issues.
    
\textit{User Engagement} was quantified through VR system data, including conversation logs with the chatbot and in-game performance metrics such as target accuracy, attempt frequency, and total playtime, comparing engagement levels between the ChatGPT and predefined chatbot.
    
\textit{Cultural heritage learning outcomes} were measured using pre- and post-assessment surveys and tests to gauge participants' understanding of the connection between cultural heritage, traditional outdoor sports, and the impact of climate change.

Additionally, we measured cognitive load using the NASA TLX and explored how participants' learning preferences, categorized by the LSQ~\cite{honey1986manual}, affected outcomes. The LSQ identifies four learner types: Activists, Reflectors, Theorists, and Pragmatists. Examining these learning styles alongside cognitive load provided insights for designing future cultural heritage applications to enhance user-centric experiences.

\subsection{Analyses}
We applied two statistical tests to evaluate usability, user engagement, learning outcomes, and cognitive load. The Mann-Whitney U test was used for usability, user engagement, and cognitive load due to its suitability for comparing independent samples without assuming a normal distribution~\cite{Gibbons_1991}. The sample t-test was used to analyze learning outcomes, which was appropriate for data meeting normal distribution criteria. This approach enabled a clear assessment of the impact of the ChatGPT-enhanced chatbot versus a predefined chatbot in a VR environment across various evaluation dimensions.

\section{Results}
Using our user study data, in this section, we report our results measured by the following dependent variables: Usability, User Engagement and Cultural heritage Learning Outcomes.

\subsection{Usability Metrics}
\begin{figure*}[ht]
  \centering
  \includegraphics[width=0.9\linewidth]{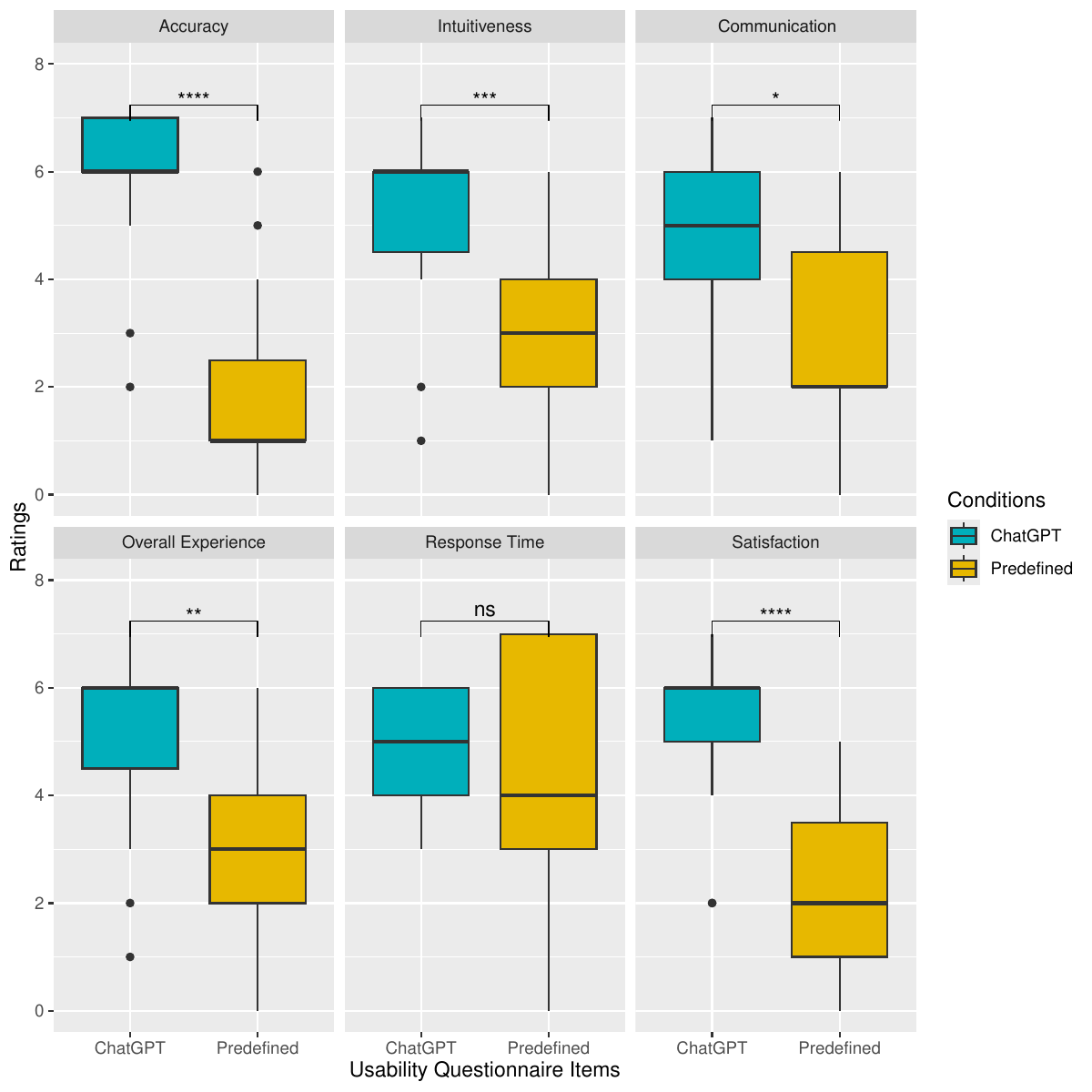}
   \caption{Usability evaluation across multiple dimensions: Comparing ChatGPT and predefined chatbot scores. *, **, ***, and **** correspond to the significance levels of \( p < .05\), \(p < .01\), \( p < .001\), and  \( p < .0001\) respectively. Response options used a Likert scale, spanning from 1 (Not at all agree) to 7 (Very much agree).}
    \label{fig:usability_comparison}
\end{figure*}

We first performed the Shapiro-Wilk test to verify the normality assumption. For non-normal distributions, the Mann-Whitney U test was used. A two-tailed Mann-Whitney U test revealed significant differences between the CGPT and PD conditions across several usability items. The results are shown in~\cref{fig:usability_comparison}.

For \textit{accuracy}, CGPT (\(M = 5.89\), \(SD = 1.37\)) was rated significantly higher than PD (\(M = 2.12\), \(SD = 1.58\)), \(U = 17.5\), \(n1 = 19\), \(n2 = 17\), \(p < 0.001\), with a large effect size (\(r = 0.761\)). Similarly, \textit{communication} scores were significantly better for CGPT (\(M = 4.95\), \(SD = 1.78\)) compared to PD (\(M = 3.18\), \(SD = 1.63\)), \(U = 78.5\), \(n1 = 19\), \(n2 = 17\), \(p = 0.008\), with a moderate effect size (\(r = 0.438\)). For \textit{intuitiveness}, CGPT (\(M = 5.32\), \(SD = 1.63\)) was scored higher than PD (\(M = 1.82\), \(SD = 1.19\)), \(U = 23\), \(n1 = 19\), \(n2 = 17\), \(p < 0.001\), with a large effect size (\(r = 0.731\)). \textit{Overall experience} was rated higher for CGPT (\(M = 5.05\), \(SD = 1.68\)) than for PD (\(M = 3.24\), \(SD = 1.35\)), \(U = 63.5\), \(n1 = 19\), \(n2 = 17\), \(p = 0.002\), with a moderate effect size (\(r = 0.518\)). There was no significant difference in \textit{response time} between CGPT (\(M = 4.68\), \(SD = 1.16\)) and PD (\(M = 4.88\), \(SD = 1.90\)), \(U = 148\), \(n1 = 19\), \(n2 = 17\), \(p = 0.675\). This suggests that response time may not be a distinguishing factor between these conditions. Finally, \textit{satisfaction} was significantly higher for CGPT (\(M = 5.37\), \(SD = 1.46\)) compared to PD (\(M = 2.47\), \(SD = 1.46\)), \(U = 28\), \(n1 = 19\), \(n2 = 17\), \(p < 0.001\), with a large effect size (\(r = 0.710\)).

\subsection{User Engagement}
The results for user engagement come from two data sources: conversation logs and in-game metrics. Detailed results can be found in appendix~\cref{fig:prompt_boxplot} and~\cref{fig:performancePlot}. A two-tailed Mann-Whitney U test was conducted to compare user engagement between the CGPT and PD conditions. As shown in appendix~\cref{fig:prompt_boxplot}, the test revealed a significant difference in prompting frequency between the two groups with \(U = 254\), \(n1 = 19\), \(n2 = 17\), \(p = 0.0034\), and a moderate effect size (\(r = 0.489\)). The results suggest that the CGPT group (\(M = 13.05\), \(SD = 5.43\)) had significantly higher prompting frequency compared to the PD group (\(M = 9.00\), \(SD = 5.66\)).

Further user engagement metrics were assessed with in-game performance data, with results detailed in~\cref{fig:performancePlot}. The two-tailed Mann-Whitney U test indicated no significant difference in the number of stones hit between the CGPT (\(M = 1.68\), \(SD = 1.73\)) and PD groups (\(M = 1.06\), \(SD = 0.83\)), \(U = 187\), \(n1 = 19\), \(n2 = 17\), \(p = 0.406\), \(r = 0.135\). Similarly, there was no significant difference in shot attempts between the CGPT (\(M = 2.58\), \(SD = 1.84\)) and PD groups (\(M = 2.41\), \(SD = 1.00\)), \(U = 155\), \(n1 = 19\), \(n2 = 17\), \(p = 0.842\), \(r = -0.034\).

However, a significant difference was found in total playtime between the two groups with \(U = 249\), \(n1 = 19\), \(n2 = 17\), \(p = 0.006\), with a medium to large effect size (\(r = 0.462\)). The CGPT group (\(M = 591.86\), \(SD = 207.22\)) engaged for a longer duration compared to the PD group (\(M = 416.28\), \(SD = 127.58\)). This increased playtime suggests that the CGPT group had significantly higher engagement levels than the PD group.

\begin{figure*}[t]
  \centering
  \includegraphics[width=\textwidth]{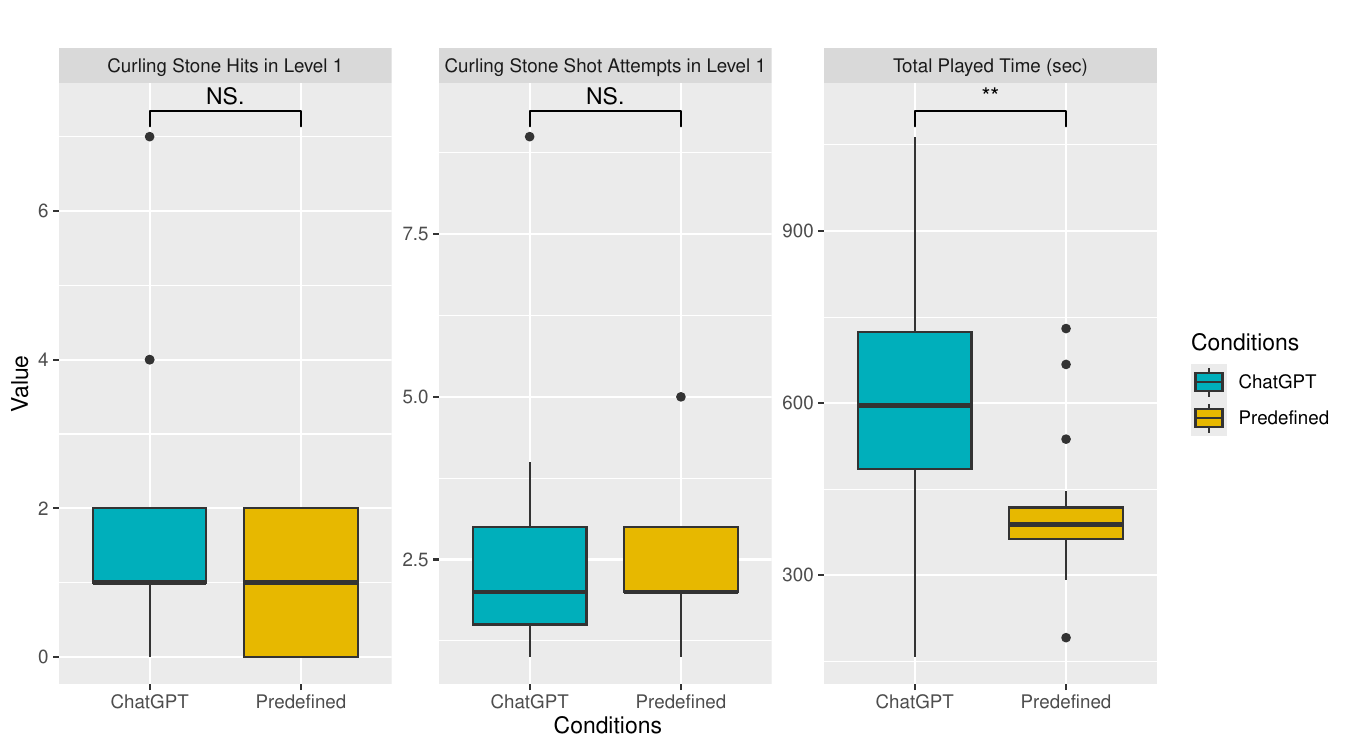}
  \caption{Comparison of User Engagement Between CGPT and PD Conditions. All times are in seconds. Asterisks denote levels of significance where *, **, and *** correspond to the significance levels of \( p < .05\), \(p < .01\), and \( p < .001\) respectively. `NS' represents no statistical significance.}
  \label{fig:performancePlot}
\end{figure*}

\subsection{Learning Outcomes}
Learning outcomes were evaluated by comparing changes in participants' knowledge before and after the VR experience under two conditions: the CGPT and the PD chatbot condition. The focus was on understanding the impact of climate change on traditional outdoor sports and the role of cultural heritage in their preservation. Sample t-tests determined the statistical significance of the changes in familiarity scores.

As depicted in appendix~\cref{fig:knowledge_gain}, the mean scores represent the change in knowledge levels after the VR experience. For the impact of climate change on traditional sports, the CGPT condition showed a mean increase of (\(M = 3.32\), \(SD = 1.83\)) compared to the PD condition (\(M = 3.06\), \(SD = 1.95\)), but this difference was not statistically significant, \(t(34) = 0.41\), \(p = 0.69\).

Regarding the role of cultural heritage in preserving traditional sports, the CGPT condition had a mean decrease (\(M = 1.05\), \(SD = 1.72\)), while the PD condition showed a mean difference of (\(M = 1.41\), \(SD = 1.18\)), with no significant difference, \(t(34) = -0.72\), \(p = 0.47\).

To further probe the retention of cultural heritage concepts, our questionnaire included four multiple-choice questions:
\begin{enumerate}
    \item What sport is associated with the Bonspiel or the ``Grand Match''?
    \item Traditionally, where were the Scottish Bonspiel games held?
    \item Why has the frequency of Bonspiel games on frozen lochs diminished?
    \item How has Scottish curling adapted in response to contemporary climate challenges?
\end{enumerate}

These questions were answered after participants finished the VR experience to assess specific knowledge without the ambiguity of open-ended responses. Although there was no statistically significant variation in responses, nearly all participants answered correctly, with only one incorrect answer.

\subsection{Cognitive Load Assessment}
Cognitive load was evaluated using the NASA TLX across six dimensions: Mental, Physical, Temporal, Performance, Effort, and Frustration, along with an Overall Workload score under CGPT and PD conditions. The results are detailed in appendix~\cref{fig:tlx_boxplot}.

Two-tailed Mann-Whitney U tests showed no significant differences between the ChatGPT (\(n1 = 19\)) and Predefined (\(n2 = 17\)) conditions for Mental (\(U = 194.0\), \(p = 0.306\)), Physical (\(U = 172.5\), \(p = 0.732\)), Temporal (\(U = 193.5\), \(p = 0.314\)), Frustration (\(U = 172.0\), \(p = 0.743\)), and Overall Workload (\(U = 159.0\), \(p = 0.949\)). Trends towards significance were observed in Performance (\(U = 110.5\), \(p = 0.108\)) and Effort (\(U = 107.5\), \(p = 0.088\)), with lower scores in the ChatGPT condition. While the results were not statistically significant, they suggest nuanced distinctions in the cognitive load experienced by participants, hinting at differential cognitive demands imposed by the two types of chatbot interactions in the VR setting.

\section{Discussion}
The discussion explores the implications of our findings for cultural heritage preservation and technology design. By comparing CGPT and PD chatbots in a VR setting, we gain insights into the potential of LLM to enhance engagement with ICH. Our discussion focuses on three key aspects of emerging technologies in safeguarding cultural traditions amid climate change.

\subsection{Enhancing Usability and Scalability in Cultural Heritage Applications: The Capabilities of LLM}

We developed a user-centric cultural heritage application, validated through a customized usability survey. The results, detailed in~\cref{fig:usability_comparison}, show significant enhancements in usability with the LLM-assisted VR experience, confirming our~\textbf{RQ1} and demonstrating LLM's adaptive and conversational strengths~\cite{Xu2023-za, Suhadi2016, izumi2024response, qi-etal-2023-art}. LLMs effectively handle diverse inquiries, foster creative thinking, and manage user expectations, surpassing the limitations of traditional predefined chatbots~\cite{Glass2001}.

While results show that LLM chatbots generally receive high accuracy ratings, these should be interpreted with caution due to the risk of LLMs generating misleading responses~\cite{augenstein2023factuality}. This phenomenon aligns with previous research on LLMs, which suggests that a human-like conversational tone may enhance clarity but does not necessarily improve accuracy~\cite{huschens2023trust, chen2023llmgenerated}. 

Aligned with previous research, LLMs can streamline technical development and facilitate scalable deployment across digital platforms~\cite{Bansal2024}. However, feedback from open-ended questions revealed technical issues such as~\textit{``Sometimes voice not recognized correctly''} and~\textit{``Experienced internet connection problems''}. These challenges highlight the dependency of LLM effectiveness on robust internet service, which may be lacking at some cultural heritage sites.

Our study affirms the role of emerging technologies like LLM in enhancing the usability of VR applications for cultural heritage, advocating their use in both preserving and educating about ICH, thus enriching the cultural heritage experience.

\subsection{Designing Future Cultural Heritage Applications: Leveraging LLM and VR to Engage Younger Audiences}
Our analysis, detailed in Appendix~\cref{fig:prompt_boxplot} and~\cref{fig:performancePlot}, reveals that users engage more frequently and for longer durations with the Scottish Bonspiel VR experience when assisted by LLMs rather than predefined chatbots. This finding, addressing \textbf{RQ2}, highlights the effectiveness of integrating LLMs' dynamic conversational capabilities with VR’s immersive visual and auditory modalities, significantly enriching the user experience.

We developed a prototype offering multimodal interaction in VR, enabling users to converse with a chatbot using their voice. We found that combining VR with LLMs enables a more interactive way to disseminate ICH knowledge and engage younger audiences. Situating this research within the framework of ``Experimental Preservation'' offers an alternative approach to safeguarding cultural knowledge, recognizing culture as a living, evolving phenomenon rather than a static object. Unlike the one-sided narration often found in cultural heritage representations~\cite{taylor2017}, our approach demonstrates that leveraging LLMs in VR empowers users to actively participate in constructing the narrative. This involves engaging in conversations about the history and development of Scottish curling against the backdrop of climate change, as illustrated in Appendix~\cref{fig:prompt_boxplot}. This is crucial for fostering critical thinking and self-awareness among younger generations toward diverse cultural heritages around the world~\cite{su12208640}.

\subsection{LLM and VR as New Learning Paradigms for Cultural Heritage Education}
From the results in~\cref{fig:knowledge_gain}, no significant differences were found in learning outcomes between CGPT and PD chatbots, this is likely due to the small sample size and limited evaluation tools. However, both groups demonstrated an increased understanding of the impact of climate change and the importance of preserving cultural heritage, addressing \textbf{RQ3}. This suggests that integrating LLM and VR can provide an alternative approach to imparting cultural heritage knowledge, particularly when on-site learning about specific cultural heritage is not possible.

\subsection{Exploring Cognitive Load and Learning Styles}
Our exploratory analysis examines the relationship between cognitive load, learning styles, and user interaction in VR-enhanced chatbot environments. Results from the NASA TLX, detailed in appendix~\cref{fig:tlx_boxplot}, show that participants engaging with ChatGPT experienced higher cognitive load, though these differences were not statistically significant, suggesting that ChatGPT's detailed responses may increase cognitive demands in a VR setting, which warrants further investigation. For learning styles, word cloud analysis revealed distinct interaction patterns between participants and ChatGPT, as depicted in appendix~\cref{fig:chatGPT_participants_word_cloud} and~\cref{fig:learning_style_wordclouds}. Activists frequently used the term `know' in their queries, indicating a quest for specific information, while `game' and `play' were common among reflectors, theorists, and pragmatists, indicating a more experiential approach. These observations suggest the potential for personalized VR learning experiences tailored to diverse learner preferences.

\section{Limitations and Future Work}
This section outlines the limitations of our study and suggests future research directions to enhance the sustainability, interdisciplinary collaborations, and technology integration of cultural heritage applications.

Further research is necessary to evaluate the real-world impact of VR and LLM technologies on cultural heritage preservation, particularly considering the novelty bias associated with emerging technologies. As highlighted in recent studies~\cite{Miguel-Alonso2024}, the initial engagement driven by the novelty of VR and LLMs could skew early assessments of their effectiveness. Investigating how digital interventions can inspire community participation in heritage preservation aligns with UNESCO's Article 15~\cite{UNESCO_BasicText_2003}, which emphasizes the role of communities, groups, and individuals in safeguarding ICH. Longitudinal studies are crucial to determine if the proposed system not only attracts initial interest due to its innovative nature but also effectively motivates individuals to engage in activities that promote cultural heritage awareness, such as attending workshops or advocating for ICH preservation on social media, over the long term.

Collaborating with cultural heritage experts, technologists, and educators is vital for enhancing LLM-delivered content's authenticity and educational quality. Involving experts in Scottish history, AI ethics, and educational psychology ensures that VR experiences are culturally sensitive and pedagogically robust. 

Future VR designs should integrate chatbot interactions more deeply into the gaming narrative. For instance, game progression could depend on engaging in dialogues with the chatbot to solve culturally relevant puzzles or obtain essential information. This approach aims to blend entertainment with learning, enhancing user engagement and achieving more profound educational impacts.

By addressing these areas, future research can build on the insights offered by this study, fostering more sustainable, engaging, and educationally rich cultural heritage applications leveraging emerging technologies.

\section{Conclusion}
This study demonstrates the effectiveness of integrating VR and LLMs technologies in enhancing the preservation and understanding of intangible cultural heritage through the ``Scottish Bonspiel VR''. Our findings highlight the capacity of LLM-powered chatbots to create immersive and educational experiences, significantly improving usability and engagement over traditional predefined chatbots.

Our research contributes to the DH community by showcasing the potential of VR and LLMs technologies to enhance accessibility and preservation of cultural heritage. This adaptability is crucial for addressing global challenges in cultural education and preserving heritage amidst environmental changes.

Future research will refine methodologies and expand the scope to encompass a broader array of cultural contexts. We will continue to explore how emerging technologies can aid in safeguarding and revitalizing heritage. This work underscores the importance of interdisciplinary collaboration in creating authentic and engaging heritage experiences, leveraging technology for both educational and conservation purposes. Our ongoing efforts aim to reignite interest in cultural practices at risk of being forgotten, ensuring they remain a vibrant part of our shared heritage.

\section*{Acknowledgements}
The authors would like to thank Joscha von Andrian for developing the Scottish Bonspiel VR experience.

%
%
\bibliographystyle{splncs04}
\bibliography{main}

\newpage
\appendix
\section{Appendix}

\begin{table}[ht]
  \caption{Characteristics of Participants.}
  \label{tbl_1}
  \centering
  \begin{tabular}{@{}p{5cm}p{2cm}p{2cm}@{}}
    \toprule
    \textbf{Characteristics} & \textbf{Count (N=36)} & \textbf{Percentage} \\
    \midrule
    \multicolumn{3}{@{}l@{}}{\textbf{Previous Experience with VR Technology:}} \\
    \quad Yes, frequently & 3 & 8.33\% \\
    \quad Yes, occasionally & 24 & 66.67\% \\
    \quad No, never & 9 & 25.00\% \\
    \midrule
    \multicolumn{3}{@{}l@{}}{\textbf{Previous Experience with Computer Games:}} \\
    \quad Highly experienced & 9 & 25.00\% \\
    \quad Moderately experienced & 8 & 22.22\% \\
    \quad Somewhat experienced & 14 & 38.89\% \\
    \quad Not experienced & 5 & 13.89\% \\
    \midrule
    \multicolumn{2}{@{}l@{}}{\textbf{Previous Awareness of Scottish Curling:}} \\ 
    \quad Aware & 13 & 36.11\% \\
    \midrule
    \multicolumn{3}{@{}l@{}}{\textbf{Learner Style:}} \\
    \quad Reflector & 22 & 61.11\% \\
    \quad Theorist & 3 & 8.33\% \\
    \quad Activist & 2 & 5.56\% \\
    \quad Both Activist and Pragmatist & 2 & 5.56\% \\
    \quad Both Reflector and Theorist & 2 & 5.56\% \\
    \quad Both Reflector and Pragmatist & 2 & 5.56\% \\
    \quad Both Theorist and Pragmatist & 2 & 5.56\% \\
    \quad Both Theorist and Reflector & 1 & 2.78\% \\
    \bottomrule
  \end{tabular}
\end{table}

\begin{table}[b]
  \caption{Dialogues designed for level 1, chatbot acts as an (Educator). }
  \label{tbl_2}
  \centering
  \begin{tabular}{@{}clp{8cm}@{}}
    \toprule
    \textbf{Items} & \textbf{Type} & \textbf{Content} \\
    \midrule
    1 & History & The Bonspiel, also known as the `Grand Match,' is a tradition that originated in Scotland. We traditionally played the Scottish Bonspiel on frozen lochs. \\
    \midrule
    2 & Climate Change & Oh no, the ice is melting. We better get inside. Warmer winters due to climate change prevent our lochs from freezing adequately for bonspiel. \\
    \midrule
    3 & Modern Adaptations & We've adapted by moving our bonspiel games indoors to artificial ice rinks. Do you know how this change has influenced the game? \\
    \midrule
    4 & Modern Adaptations & Moving indoors can lead to more controlled ice conditions, precise shot-making, and a comfortable fan experience, while potentially altering the traditional outdoor ambiance of the game. You can shoot the curling stone now. \\
    \bottomrule
  \end{tabular}
\end{table}

\begin{table}[b]
  \caption{Dialogues designed for level 2, chatbot acts as an (Evaluator) for participants' responses.}
  \label{tbl_3}
  \centering
  \begin{tabularx}{\textwidth}{@{}cX X X@{}}
    \toprule
    \textbf{Items} & \textbf{Type} & \textbf{Questions} & \textbf{Answers} \\
    \midrule
    5 & History & Can you tell me where traditional Scottish Bonspiel games were held? & Traditional Scottish Bonspiel games were played on frozen lakes, offering a unique, challenging environment that artificial ice can't fully replicate. \\
    \midrule
    6 & Climate Change & Do you remember why we cannot play bonspiel on our lochs anymore? & Climate change and warmer winters prevent our lochs from freezing adequately. This has significantly changed the sport, shifting it from its outdoor roots to indoor venues. \\
    \midrule
    7 & Modern Adaptations & How has Scottish curling adapted to modern climate conditions? & We have moved our games indoors to artificial ice rinks. This allows the sport to continue and evolve, including the development of modern equipment and strategies. \\
    \bottomrule
  \end{tabularx}
\end{table}


\begin{figure}[ht]
  \centering
  \begin{subfigure}[b]{0.48\linewidth}
    \includegraphics[width=\linewidth,keepaspectratio]{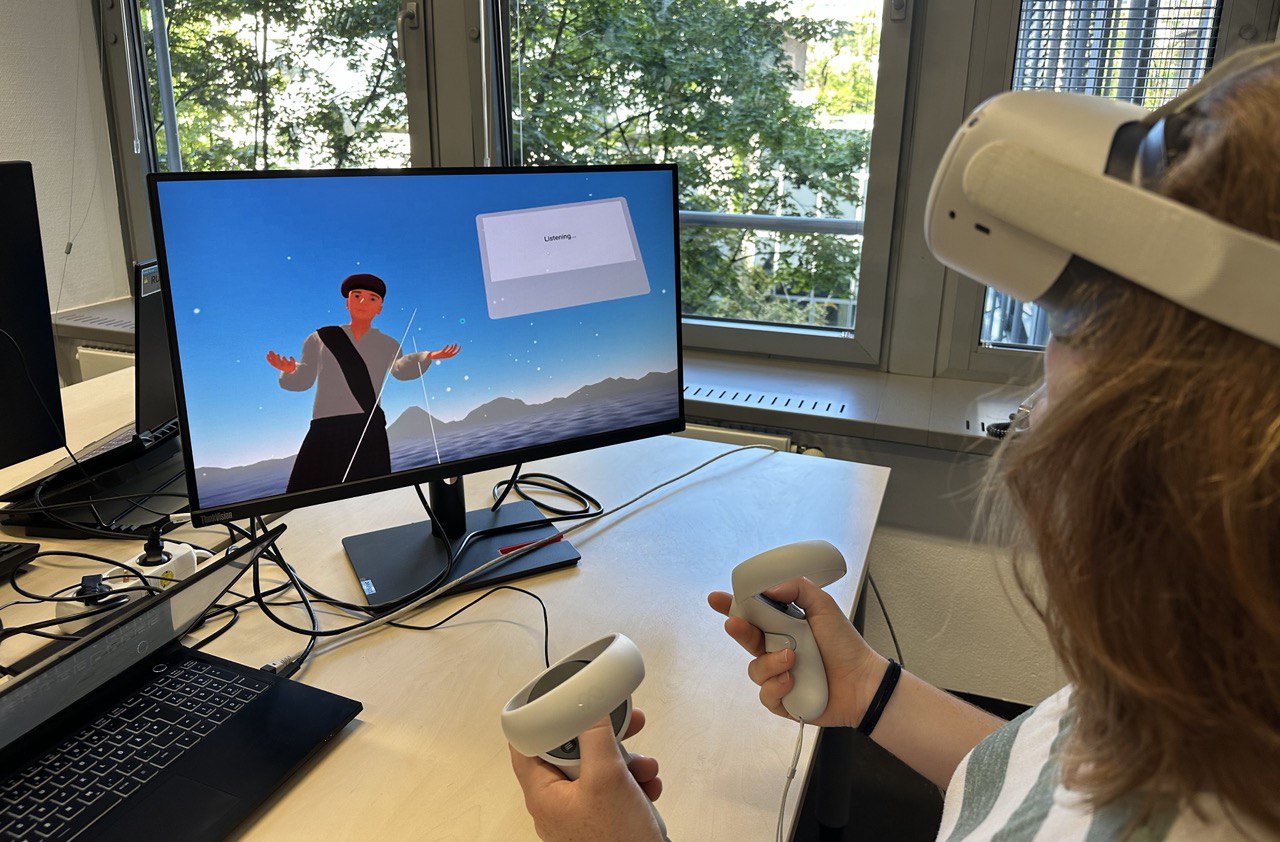}
    \caption{Players engage in conversations with the cultural heritage ambassador.}
    \label{fig:fig_player_a}
  \end{subfigure}
  \hfill
  \begin{subfigure}[b]{0.48\linewidth}
    \includegraphics[width=\linewidth,keepaspectratio]{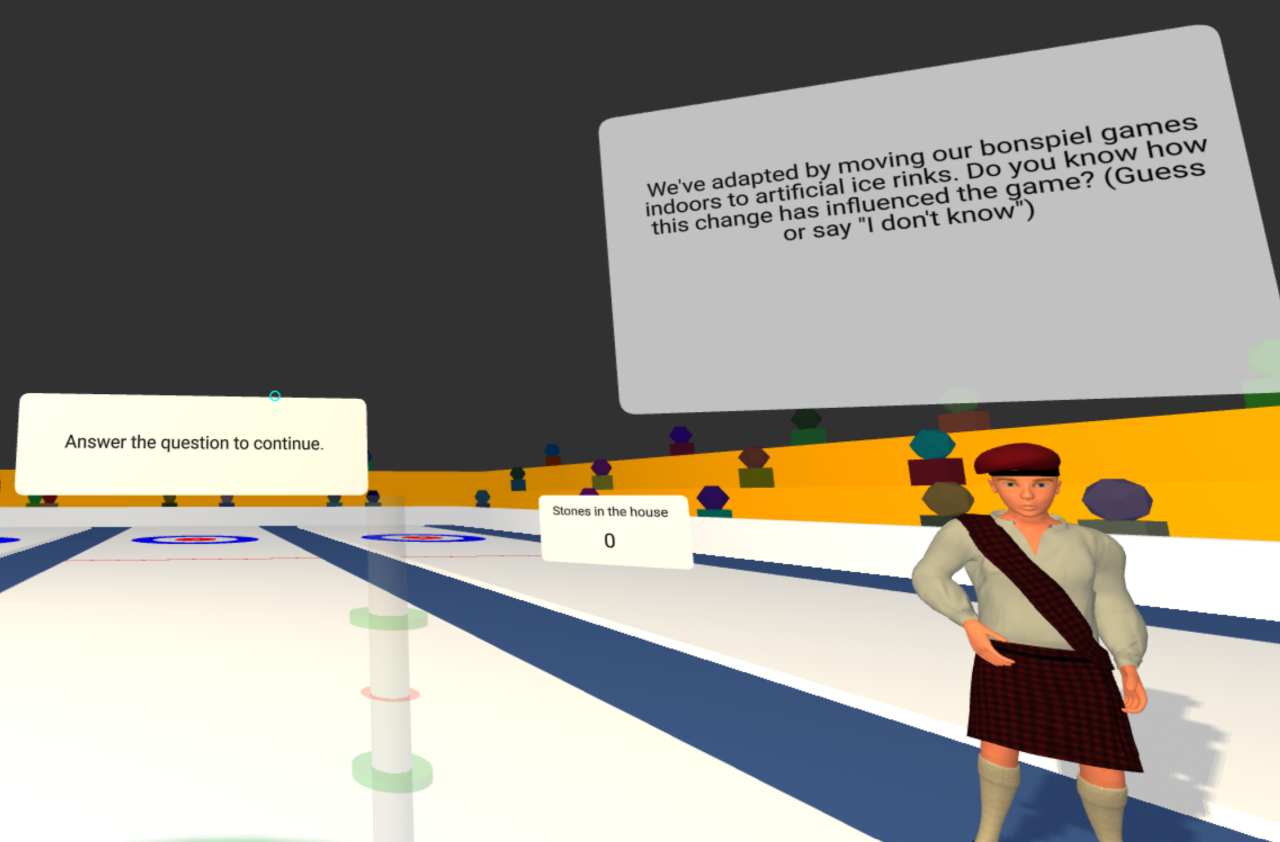}
    \caption{The game transitions from a frozen lake to an indoor ice rink.}
    \label{fig:fig_player_b}
  \end{subfigure}
  \caption{The design of the game environment for Level 1 and Level 2.}
  \label{fig:fig_player}
\end{figure}

\begin{figure*}[b]
  \centering
  \includegraphics[width=\linewidth]{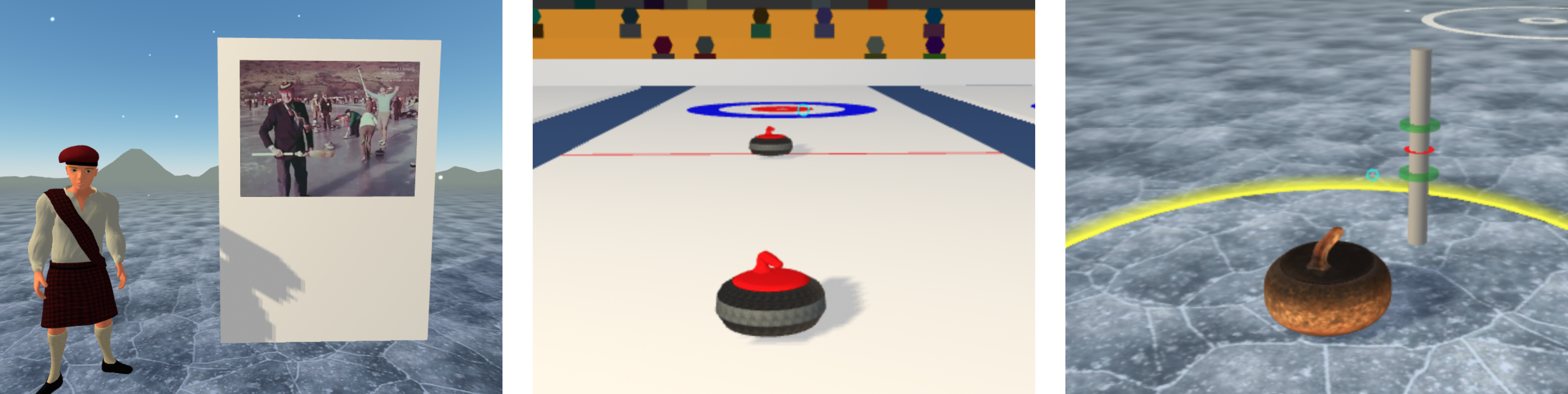}
  \caption{Objects in the VR scene are inspired by Scottish curling heritage artifacts.}
  \label{fig_all_artifacts}
\end{figure*}


\begin{figure*}[b]
  \centering
  \includegraphics[width=0.9\textwidth]{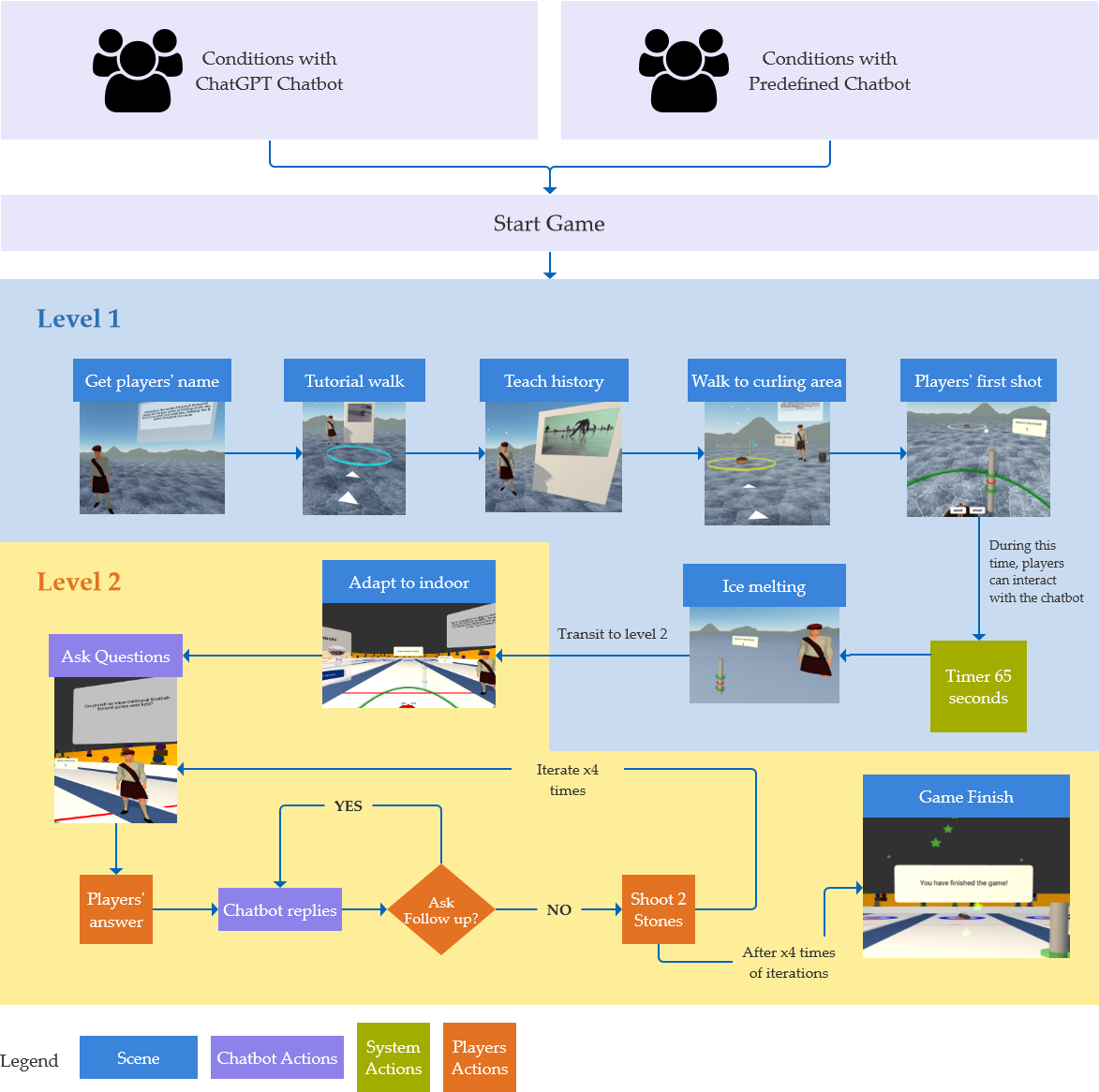}
  \caption{The overview of our procedure and experimental setup.}
  \label{fig_2}
\end{figure*}

\begin{table}[b]
  \caption{Overview of usability metrics measured.}
  \label{tbl_usability}
  \centering
  \begin{tabularx}{\textwidth}{@{}p{0.45\textwidth}Xc@{}}
    \toprule
    \textbf{Usability Questions} & \textbf{Metrics} & \textbf{Likert Scale (1-7)} \\
    \midrule
    Did the chatbot provide relevant and accurate responses? & Accuracy & Poor to Excellent \\
    \midrule
    How intuitive was the chatbot in responding to your queries? & Intuitiveness & Not at all to Extremely \\
    \midrule
    Did you encounter any difficulties in conveying your message? & Communication & Very Difficult to Not Difficult \\
    \midrule
    How would you rate your overall experience with the chatbot? & Overall Experience & Poor to Excellent \\
    \midrule
    How would you rate the response time of the chatbot? & Response Time & Too Slow to Very Fast \\
    \midrule
    How satisfied were you with the response quality? & Satisfaction & Very Dissatisfied to Very Satisfied \\
    \bottomrule
  \end{tabularx}
\end{table}



\begin{figure*}[b]
  \centering
  \includegraphics[width=0.8\textwidth]{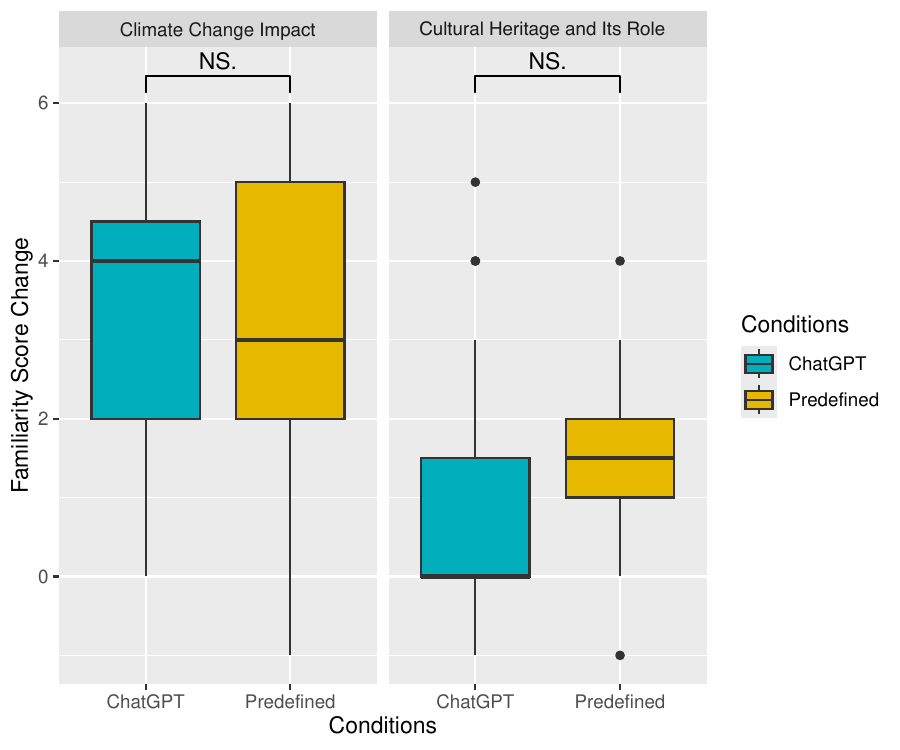}
  \caption{Comparison of Familiarity Score Changes Before and After VR Experience Between CGPT and PD Chatbots Regarding the Impact of Climate Change on Traditional Outdoor Sports and the Role of Cultural Heritage in Traditional Sports Preservation. `NS' represents no statistical significance.}
  \label{fig:knowledge_gain}
\end{figure*}

\begin{figure*}[b]
  \centering
  \includegraphics[width=0.8\textwidth]{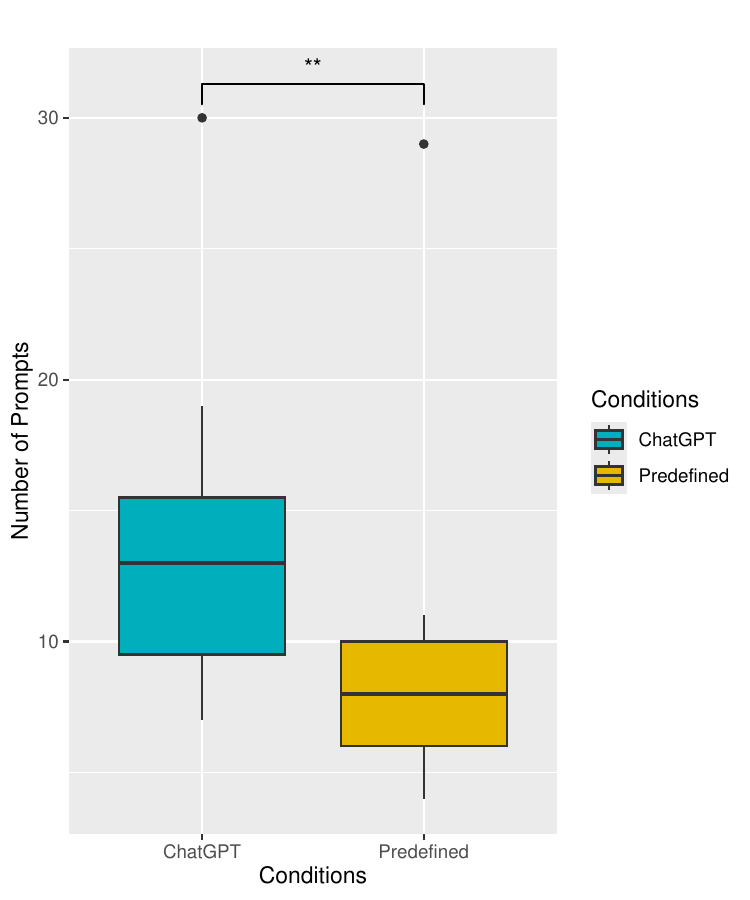}
  \caption{Comparison of User Frequency of Prompts within CGPT and PD conditions. Asterisks denote levels of significance where *, **, and *** correspond to \( p < .05\), \( p < .01\), and \( p < .001\) respectively.}
  \label{fig:prompt_boxplot}
\end{figure*}

\begin{figure*}[b]
  \centering
  \includegraphics[width=\linewidth]{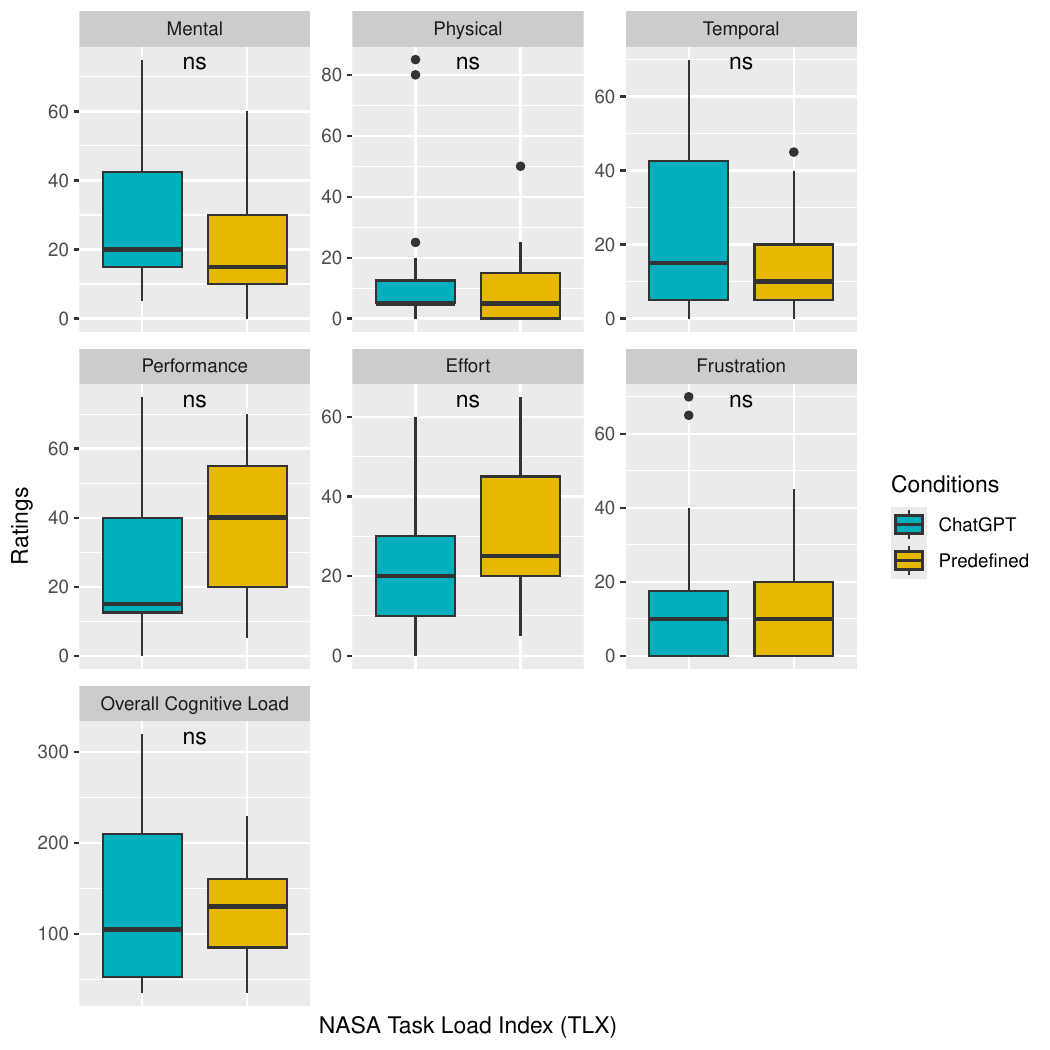}
   \caption{Comparison of workload ratings using NASA TLX between ChatGPT and predefined conditions. `NS' represents no statistical significance.}
    \label{fig:tlx_boxplot}
\end{figure*}

\begin{figure*}[b]
  \centering
  \begin{subfigure}[b]{0.48\linewidth}
    \includegraphics[width=\linewidth,keepaspectratio]{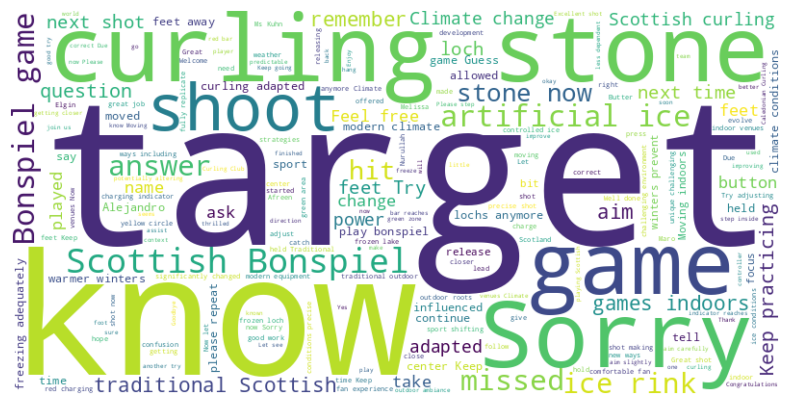}
    \caption{ChatGPT's responses.}
    \label{fig:chatGPT_responses}
  \end{subfigure}
  \hfill
  \begin{subfigure}[b]{0.48\linewidth}
    \includegraphics[width=\linewidth,keepaspectratio]{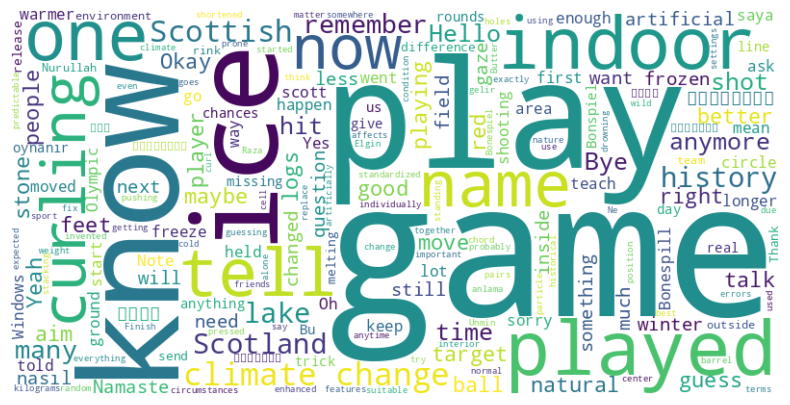}
    \caption{Participants' prompts.}
    \label{fig:participants_prompts}
  \end{subfigure}
  \caption{Word cloud analysis of ChatGPT and participant prompt interactions.}
  \label{fig:chatGPT_participants_word_cloud}
\end{figure*}

\begin{figure*}[b]
  \centering
  \begin{subfigure}[b]{0.48\linewidth}
    \includegraphics[width=\linewidth,keepaspectratio]{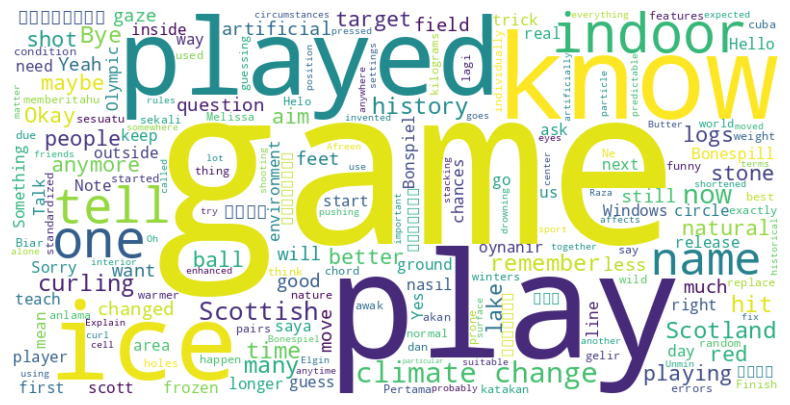}
    \caption{Reflector learners' prompt, $n=178$.}
    \label{fig:reflector_prompt}
  \end{subfigure}
  \hfill
  \begin{subfigure}[b]{0.48\linewidth}
    \includegraphics[width=\linewidth,keepaspectratio]{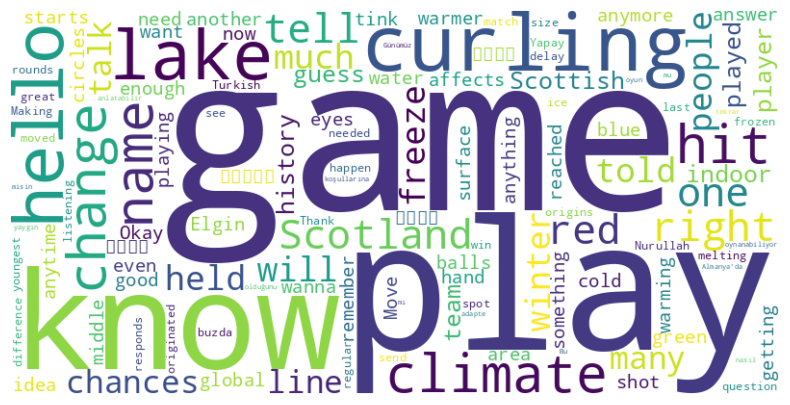}
    \caption{Theorist learners' prompt, $n=60$.}
    \label{fig:theorist_prompt}
  \end{subfigure}
  \hfill
  \begin{subfigure}[b]{0.48\linewidth}
    \includegraphics[width=\linewidth,keepaspectratio]{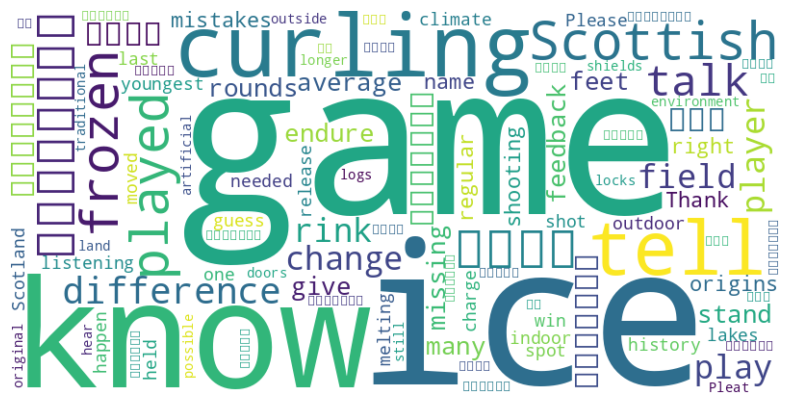}
    \caption{Pragmatist learners' prompt, $n=39$.}
    \label{fig:pragmatist_prompt}
  \end{subfigure}
  \hfill
  \begin{subfigure}[b]{0.48\linewidth}
    \includegraphics[width=\linewidth,keepaspectratio]{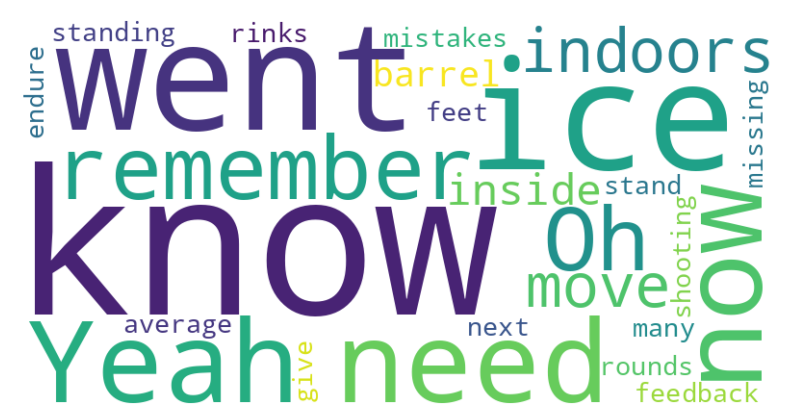}
    \caption{Activist learners' prompts, $n=14$.}
    \label{fig:activist_prompt}
  \end{subfigure}
  \caption{Word cloud analysis of participant prompts by learning style.}
  \label{fig:learning_style_wordclouds}
\end{figure*}

\end{document}